\documentclass[conference,a4paper,10pt]{IEEEtran}
 \usepackage{mathtools}
 \usepackage{amsfonts}
 \usepackage{amsmath}
 \usepackage{bbm}
 \usepackage{amsthm}
 \usepackage{amssymb}

\usepackage{tablefootnote}
\usepackage{url}

\usepackage{algorithm}
\usepackage{algorithmicx}
\usepackage{algpseudocode}
\usepackage{booktabs}
\usepackage{textcomp}
\usepackage{enumitem}
\usepackage{graphicx}
\usepackage{subcaption}
\usepackage{graphicx,epstopdf}
\usepackage{url}
\usepackage{cite}
\usepackage{soul}
\usepackage{balance}
\usepackage{multirow}

\usepackage{tkz-kiviat}
\usetikzlibrary{arrows}
\usetikzlibrary{matrix}
\usepackage{placeins}

\usepackage[dvipsnames, table]{xcolor}
\definecolor{lightgray}{gray}{0.9}

\newcommand{\sofie}[1]{%
    \ifcomments
        {\color{magenta}[Sofie: #1]}%
    \fi
}

\newcommand{\emiel}[1]{%
  \ifcomments
    {\color{red}[Emiel: #1]}%
  \fi
}

\newif\ifcomments
\commentstrue      
\commentsfalse   

\ifCLASSINFOpdf
\else
\fi
\begin{document}

%
\title{Frequency-Adaptive Multi-Band Architecture for Upper Mid-Band MIMO Systems}
\author{\IEEEauthorblockN{Emiel Vanspranghels\IEEEauthorrefmark{1}, Zhuangzhuang~Cui\IEEEauthorrefmark{1}, and Sofie Pollin\IEEEauthorrefmark{1}\IEEEauthorrefmark{2}}
\IEEEauthorblockA{
\IEEEauthorrefmark{1}WaveCoRE, Department of Electrical Engineering (ESAT), KU Leuven, Belgium\\
\IEEEauthorrefmark{2}Interuniversity Microelectronics Centre (imec), Leuven, Belgium\\}
Email: \texttt{\{firstname.lastname\}@kuleuven.be}}

%



\maketitle

\begin{abstract}
FR3 ($\approx$7–24~GHz), also referred to as the upper mid-band, has recently emerged as promising spectrum for 6G; however, its propagation and MIMO characteristics vary significantly with frequency and environment, and spectrum availability may be intermittent due to incumbents. Using site-specific ray tracing (Sionna RT) in representative indoor and outdoor scenarios, we evaluate 7, 10, 14, 20, and 24~GHz under SISO and MIMO configurations. The results show that FR3 exhibits propagation characteristics intermediate between sub-6~GHz and mmWave bands while supporting meaningful spatial multiplexing, albeit with strong site dependence. Motivated by these findings, we propose a fully digital frequency-adaptive multi-band MIMO architecture that repurposes ADCs/DACs and baseband processing resources across FR3 subbands via switching, enabling dynamic trade-offs between bandwidth (spectrum gain) and antenna consolidation (MIMO gain) under availability and channel constraints. Simulation results demonstrate that exploiting additional spectrum is often optimal, while adaptive resource repurposing becomes beneficial when subbands are unavailable or when multiplexing gains are concentrated at specific frequencies.
\end{abstract}


\section{Introduction}
Recently, the upper mid-band, also referred to as Frequency Range~3 (FR3), has emerged as a promising spectrum region offering a favorable trade-off between coverage and throughput for envisioned multi-service 6G deployments \cite{cui25dyspan}. Positioned between sub-6~GHz and millimeter-wave (mmWave) bands, FR3 enables higher data rates than conventional mid-band systems while retaining more robust propagation characteristics than mmWave communications. However, cellular operation in FR3 must coexist with incumbent services such as satellite communications and radio astronomy, which imposes stringent interference and spectrum usage constraints \cite{fr3vision}. As a result, spectrum availability and dynamic spectrum access are critical enablers of practical deployment. When static spectrum partitioning is infeasible, dynamic spectrum sharing can be supported by the spatial multiplexing and interference suppression capabilities of massive multiple-input multiple-output (MIMO), as demonstrated in prior coexistence studies \cite{digitalcoexistence}. FR3 spectrum sharing hence critically depends on both MIMO and spectrum flexibility.

To achieve peak capacity, future cellular systems are expected to exploit a wide portion of the FR3 spectrum through dynamic band selection and aggregation based on user-specific channel conditions and service requirements \cite{cellularuppermid}. However, FR3 spans several gigahertz with pronounced 
variations in propagation loss, antenna aperture, hardware efficiency, and array response, challenging conventional fixed-band MIMO architectures. While shared-aperture antenna designs allow multi-band integration, fundamental architectural questions remain regarding the number of RF frontends, up- and downconversion chains, and simultaneously active subbands. Efficient operation across FR3 therefore requires architectures that support scalable beamforming and frequency-adaptive spatial processing under practical hardware constraints.

Recent architectural studies have explored wideband FR3 operation using true-time-delay (TTD) elements to mitigate beam squint and enable frequency-invariant beamforming \cite{TTD,ttdcoexistence}. Although effective, TTD-based solutions face challenges related to hardware complexity, power consumption, and scalability in large antenna arrays. Alternative approaches, such as tri-hybrid MIMO architectures, introduce additional beamforming layers to enable frequency-selective spatial processing using reconfigurable electromagnetic structures, including dynamic metasurface antennas \cite{heath2025trihybridmimoarchitecture,dynamicmetasurface}. These designs offer improved scalability without requiring per-element TTDs.

In \cite{mizmizi2025hybridmimouppermidband}, a systematic classification of digital and hybrid MIMO architectures is presented, with emphasis on processing complexity and energy efficiency. Building on this classification, this work extends the analysis to extreme MIMO configurations and evaluates the peak and aggregated spectral efficiency of multi-band MIMO architectures using site-specific ray-tracing channels \cite{raytracingzachrudranil}.

In this paper, we propose and analyze a frequency-adaptive MIMO architecture for upper mid-band systems that enables flexible multi-band operation across FR3. The proposed approach supports dynamic frequency selection and resource repurposing based on channel conditions, coexistence requirements, and hardware constraints, while explicitly distinguishing between long-term architectural design choices and short-term operational adaptations. The main contributions are summarized as follows:
\begin{itemize}
\item We introduce a frequency-adaptive MIMO architecture framework that separates design-time hardware configuration from run-time frequency and resource allocation, enabling flexible reuse of baseband processing resources across FR3.
\item We evaluate the proposed framework through site-specific simulations, demonstrating performance gains from dynamic frequency selection and adaptive resource allocation across subbands.
\end{itemize}

The remainder of this paper is organized as follows. Section~\ref{freqadap} introduces the frequency-adaptive architecture and contrasts it with frequency-partitioned and frequency-integrated designs. Section~\ref{designrun} details design-time and run-time resource allocation, including coexistence and channel-aware adaptation. Section~\ref{simres} presents simulation results, and Section~\ref{conclusion} concludes the paper.

\section{The Frequency-Adaptive Architecture}\label{freqadap}

Multi-band MIMO architectures have been classified in \cite{mizmizi2025hybridmimouppermidband}, proposing eight categories based on three key criteria. First, architectures are distinguished by their frequency usage: frequency-integrated designs can simultaneously access all FR3 subbands, whereas frequency-partitioned designs operate on a single subband at a time. Independent operation across subbands can be enabled through either in-band full-duplex \cite{inbandfullduplex} or subband full-duplex \cite{subbandfullduplex}, thereby alleviating system-wide time division duplex (TDD) constraints. Second, the classification differentiates between fully digital and hybrid analog–digital beamforming architectures. Third, architectures are categorized based on RF chain allocation, distinguishing between designs with dedicated RF chains per subband and those that share RF chains across multiple subbands.

In this work, we introduce the concept of a frequency-adaptive architecture, which occupies an intermediate position between frequency-integrated and frequency-partitioned designs, while focusing exclusively on fully digital operation. Unlike frequency-partitioned architectures, the frequency-adaptive approach supports simultaneous operation over multiple subbands. At the same time, it differs from frequency-integrated architectures by enabling dynamic repurposing of RF chain resources, such that not all subbands must be accessed concurrently. With respect to RF chain allocation, the proposed architecture supports both dedicated and shared RF frontends depending on the subband definitions, while employing dedicated up- and downconversion stages and shared Analog-to-Digital Converters (ADCs) and Digital-to-Analog Converters (DACs).



\begin{figure*}[!t]
\centering

\begin{subfigure}{0.95\textwidth}
  \centering
  \includegraphics[width=\textwidth]{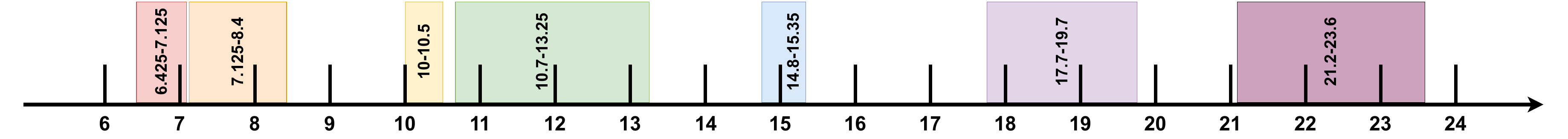}
  \caption{FR3 divided into subbands.}
  \label{frequencyrange}
\end{subfigure}

\vspace{0.4em}

\begin{subfigure}{0.95\textwidth}
  \centering
  \includegraphics[width=\textwidth]{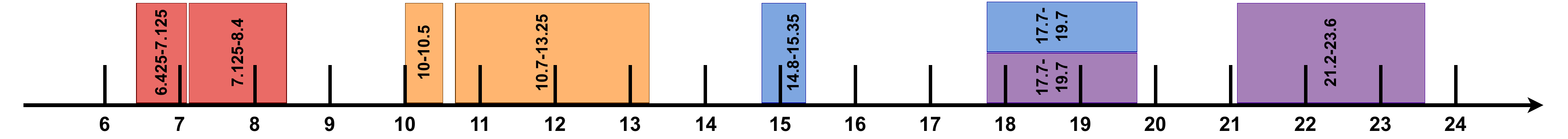}
  \caption{Hardware allocation across FR3 subbands, where overlap creates further MIMO degrees of freedom.}
  \label{hwalloc}
\end{subfigure}
\caption{FR3 subband division and hardware allocation.}
\vspace{-6pt}
\end{figure*}


Fig.~\ref{frequencyrange} shows a possible division of FR3 into subbands. Owing to hardware constraints, such as the operational fractional bandwidth of power amplifiers, RF frontends can realistically span one or several subbands, but are unlikely to cover the entire FR3 range. Site-dependent propagation statistics and user-specific channel conditions determine which subbands maximize spectral efficiency at a given time, while the optimal allocation of RF chains per subband depends on achievable MIMO gains.




\begin{figure}[!t]
  \centering

  \begin{subfigure}{0.95\columnwidth}
    \centering
    \includegraphics[width=\columnwidth]{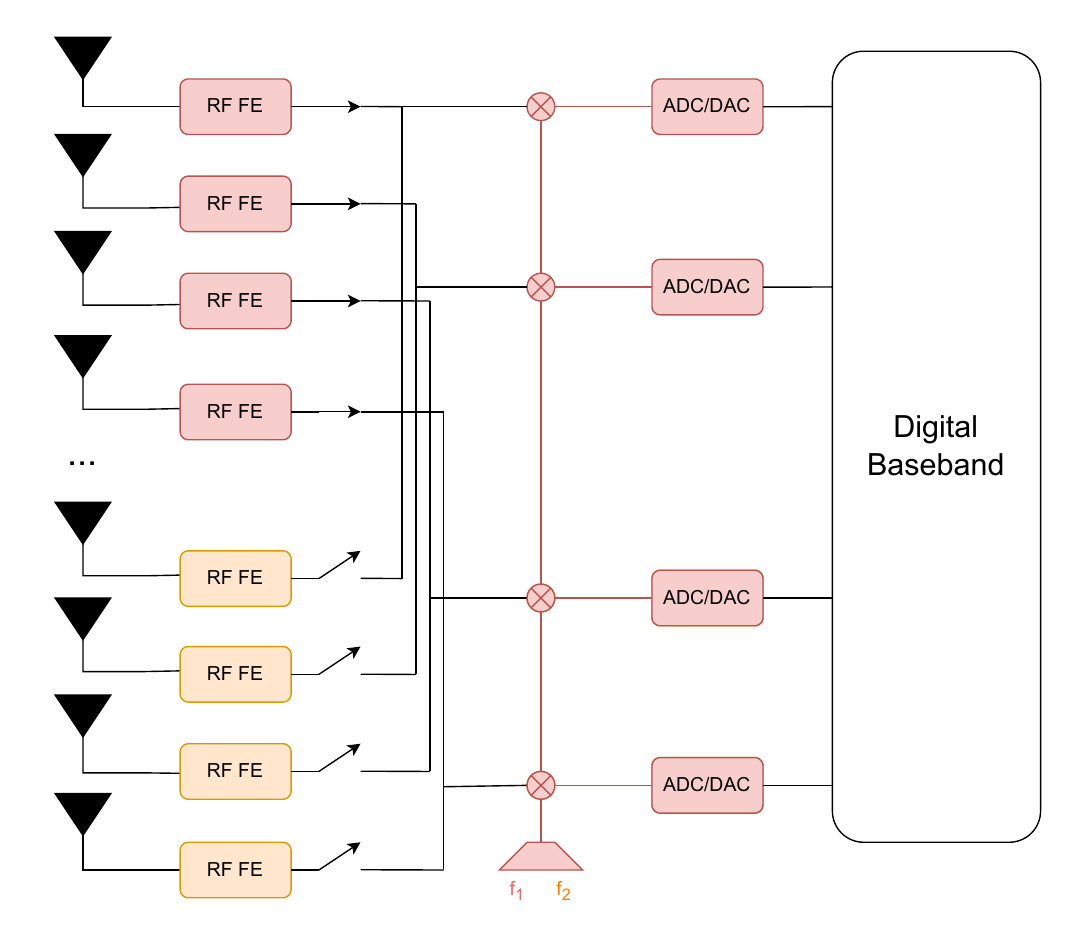}
    \vspace{-12pt}
    \caption{The fully digital frequency-partitioned architecture. The number of ADCs/DACs is equal to the number of antennas in one subband.}
    \label{freqpar_architecture}
  \end{subfigure}

  \vspace{0.4em}

  \begin{subfigure}{0.95\columnwidth}
    \centering
    \includegraphics[width=\columnwidth]{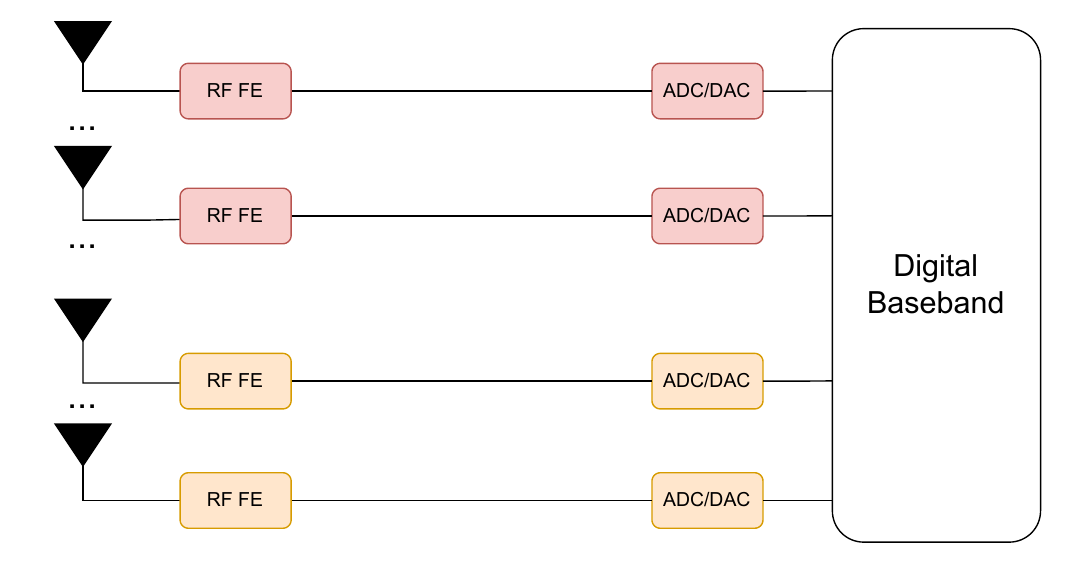}
    \vspace{-12pt}
    \caption{The fully digital frequency-integrated architecture. The number of ADCs/DACs is equal to the total number of antennas in all subbands.}
    \label{freqint_architecture}
  \end{subfigure}


  \begin{subfigure}{0.95\columnwidth}
    \centering
    \includegraphics[width=\columnwidth]{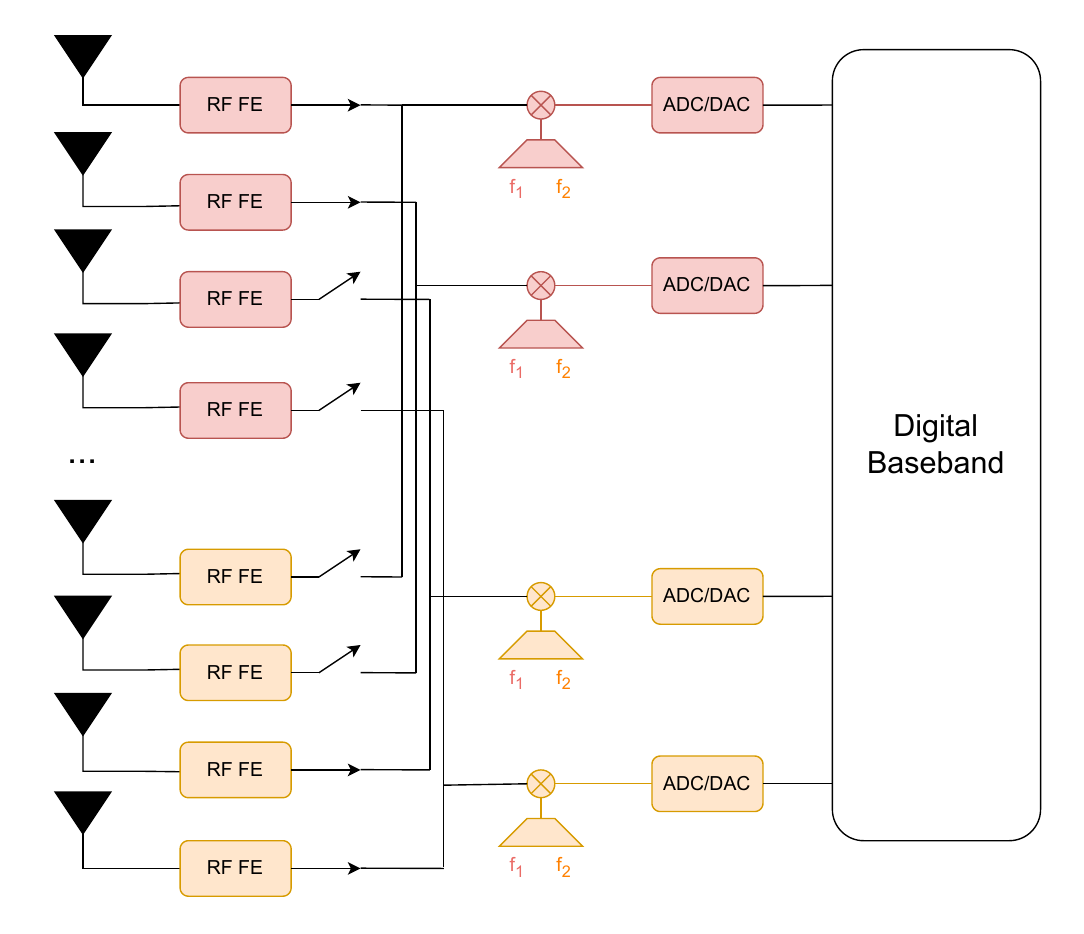}
    \vspace{-12pt}
    \caption{The fully digital frequency-adaptive architecture. The number of ADCs/DACs is equal to the number of antennas in one subband, but can be different to provide even more dynamic adaptivity.}
    \label{architecture}
  \end{subfigure}

  \caption{Fully digital multi-band MIMO architectures considered in this work.}
  \label{fig:architectures}
\end{figure}

Fig.~\ref{freqpar_architecture} illustrates a simplified two-subband frequency-partitioned architecture, in which all baseband processing resources are dedicated to a single active subband. Fig.~\ref{freqint_architecture} shows the frequency-integrated architecture, where each subband is served by fully dedicated RF chains and ADCs/DACs.

Fig.~\ref{architecture} presents an illustration of a fully digital frequency-adaptive MIMO transceiver. The core principle is to repurpose ADCs/DACs and baseband processing resources in response to channel conditions, with subband selection implemented through a switching network that connects individual ADCs/DACs to different RF frontends. 

By provisioning multiple sets of RF frontends and antennas, each designed to operate over one or multiple subbands, and interconnecting them to the available ADCs/DACs via a switching network, the architecture enables either maximization of MIMO gain within a single subband or a trade-off between MIMO gain and spectrum gain by distributing ADC/DAC and baseband processing resources across multiple subbands. A key requirement for coherent MIMO operation is phase coherence across RF frontends, which has been demonstrated in a multi-antenna FR3 transceiver \cite{FR3phasecoherence}.



\section{Design-Time and Run-Time Hardware and Frequency Allocation}\label{designrun}

Decisions regarding hardware and frequency allocation occur at both design-time and run-time. During design-time, the number of FR3 subbands and their frequency ranges must be determined. For each subband, antennas and frontends can either be dedicated or shared across multiple subbands, with reconfigurability enabling efficient hardware use. The number of antennas per set is also a design choice. These decisions can be guided by site-specific ray-tracing results while accounting for incumbent services. Key hardware constraints include the frequency compatibility of power amplifiers, tunability of local oscillators, and the supported bandwidths of ADCs/DACs. \sofie{Nice paragraph, but very generic. Get more concrete later. }

At run-time, ADCs/DACs and available baseband processing \emiel{old:and reconfigurable frontend modules} can be dynamically repurposed in response to varying conditions, such as incumbent interference and per-user frequency-dependent channel quality. By serving users across multiple frequency options, the system can balance coverage and throughput optimally. For instance, when a user moves indoors, switching to lower frequencies can reduce penetration losses by several dB, thereby improving link reliability \cite{penetrationloss}. \sofie{Make sure you have numbers to support this}

A possible design-time hardware configuration is illustrated in Fig.~\ref{hwalloc}, where each color represents a set of antennas allocated to corresponding frequency subbands. In this example, four sets of RF frontend modules are shown. Each different set $i$ consists of $N_i$ identical frontends and antennas, and is assigned to dedicated or overlapping subbands. This configuration guarantees a maximum fractional bandwidth of 29\% per set \cite{fbwPA}. At run-time, one can opt for maximal spectrum usage, or an optimization of MIMO gain, e.g. in case of limited spectrum access.

\section{Simulation Results}\label{simres}
The benefit of dynamic frequency allocation, corresponding to frequency-partitioned architectures, over fixed-frequency operation has been demonstrated in \cite{cellularuppermid}. This gain arises from the strong frequency dependence of FR3 channel characteristics, where lower frequencies exhibit reduced penetration loss and higher frequencies provide increased directionality and improved interference mitigation \cite{cellularuppermid}. 
Since channel conditions vary significantly across users, we advocate frequency-adaptive architectures rather than strictly frequency-partitioned designs. The main advantage of the frequency-adaptive approach over frequency-integrated architectures stems from its ability to repurpose ADC/DAC and baseband resources when certain subbands are unavailable.

In a first experiment, we evaluate spectral efficiency averaged over 100 user locations in an indoor laboratory environment and 20 user positions in an urban macro scenario, for 5 different FR3 frequencies. These frequencies do not perfectly align with the subbands in Fig.~\ref{frequencyrange}, but are chosen to evenly cover the whole FR3. Site-specific ray-tracing channels are used to compute the numbers reported in Tab.~\ref{tab:spectral_efficiency_lab} and Tab.~\ref{tab:spectral_efficiency_uma}.


\begin{table}[!t]
\centering
\caption{Spectral efficiency (bits/s/Hz) across FR3.}
\label{tab:spectral_efficiency}

\begin{subtable}[t]{\linewidth}
\centering
\caption{Indoor laboratory environment.}
\label{tab:spectral_efficiency_lab}
\begin{tabular}{c | c c c c c}
\hline
\textbf{MIMO size} &
\textbf{7 GHz} & \textbf{10 GHz} & \textbf{14 GHz} & \textbf{20 GHz} & \textbf{24 GHz} \\
\hline
\textbf{1x1} & 6.525  & 6.553  & \cellcolor{lightgray}6.527  & \cellcolor{lightgray}6.520  & \cellcolor{lightgray}6.451  \\
\textbf{2x2} & 9.145  & \cellcolor{lightgray}9.286  & 8.969  & 9.202  & 9.126  \\
\textbf{3x3} & 12.252 & 11.917 & 11.427 & 11.962 & 11.733 \\
\textbf{4x4} & \cellcolor{lightgray}15.617 & 15.299 & 15.045 & 15.348 & 15.141 \\
\textbf{5x5} & 17.986 & 17.460 & 17.126 & 17.359 & 17.208 \\
\textbf{6x6} & 20.486 & 19.604 & 19.278 & 19.406 & 19.182 \\
\textbf{7x7} & 23.606 & 22.576 & 22.247 & 22.579 & 21.980 \\
\textbf{8x8} & 25.738 & 24.568 & 24.125 & 24.459 & 23.813 \\
\textbf{9x9} & 28.083 & 26.606 & 26.150 & 26.394 & 25.630 \\
\hline
\end{tabular}
\end{subtable}

\vspace{0.6em}

\begin{subtable}[t]{\linewidth}
\centering
\caption{Outdoor urban macro environment.}
\label{tab:spectral_efficiency_uma}
\begin{tabular}{c | c c c c c}
\hline
\textbf{MIMO size} &
\textbf{7 GHz} & \textbf{10 GHz} & \textbf{14 GHz} & \textbf{20 GHz} & \textbf{24 GHz} \\
\hline
\textbf{1x1} & 6.302  & 6.720  & 6.258  & 6.514  & \cellcolor{lightgray}6.553  \\
\textbf{2x2} & \cellcolor{lightgray}8.737  & \cellcolor{lightgray}9.152  & \cellcolor{lightgray}8.537  & \cellcolor{lightgray}8.649  & 8.677  \\
\textbf{3x3} & 10.405 & 10.882 & 10.250 & 10.184 & 10.302 \\
\textbf{4x4} & 12.377 & 13.339 & 13.004 & 13.279 & 12.769 \\
\textbf{5x5} & 13.587 & 14.635 & 14.438 & 14.815 & 14.170 \\
\textbf{6x6} & 14.743 & 15.934 & 15.723 & 16.036 & 15.340 \\
\textbf{7x7} & 16.354 & 17.956 & 17.696 & 18.056 & 17.434 \\
\textbf{8x8} & 17.373 & 19.070 & 18.878 & 19.336 & 18.645 \\
\textbf{9x9} & 18.422 & 20.200 & 19.983 & 20.446 & 19.690 \\
\hline
\end{tabular}
\end{subtable}
\vspace{-12pt}
\end{table}

The results indicate that spectrum gain often dominates MIMO gain, favoring frequency-integrated operation when all subbands are available. However, under partial spectrum availability due to incumbent activity \cite{incumbent}, and when sufficient spatial multiplexing is possible, performance is improved by dynamically repurposing baseband resources across available subbands. 
Moreover, very large MIMO configurations exceeding 256 antenna ports have been proposed for FR3 \cite{giganticMIMO}.

The results in Tab.~\ref{tab:spectral_efficiency_lab} and Tab.~\ref{tab:spectral_efficiency_uma} assume fixed transmit power per antenna, allowing spectral efficiency to be aggregated across subbands up to a total 9×9 channel. The proposed approach exploits all available subbands to maximize spectrum gain, and subsequently distributes MIMO resources across subbands to maximize the overall sum spectral efficiency.

In the indoor scenario, the maximum sum spectral efficiency of 44.401~bits/s/Hz is achieved by allocating 4×4 MIMO at 7~GHz, 2×2 MIMO at 10~GHz, and 1×1 MIMO at the remaining frequencies under a total 9×9 antenna constraint. In the outdoor scenario, the optimum of 41.628~bits/s/Hz is obtained by allocating 1×1 MIMO at 24~GHz and 2×2 MIMO at the remaining frequencies. These results validate the site-specific nature of the design and the benefit of dynamically allocating MIMO resources. Overall, the key takeaway is to exploit as much spectrum as feasible while concentrating MIMO capabilities where they yield the largest gains. The unavailability of one or more subbands can therefore be efficiently addressed through dynamic hardware repurposing. \emiel{new table with only 7 GHz and 24 GHz bands available (for which there is no space left)}


\begin{figure}[!t]
  \centering
  {\includegraphics[width=3.3in]{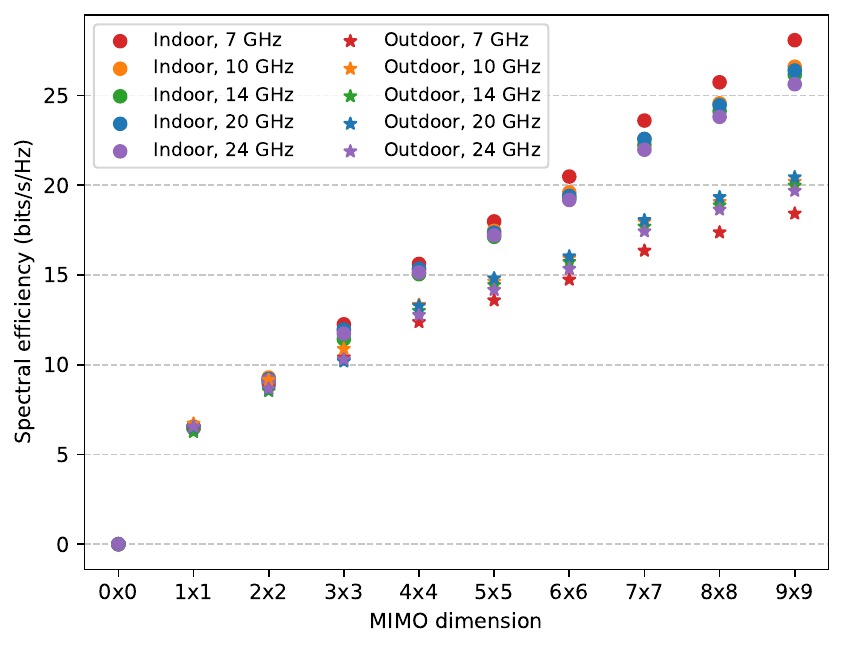}}
  \caption{Per-frequency sum spectral efficiency in function of MIMO dimensions.} 
  \label{sum_SE}
  \vspace{-12pt}
\end{figure}

Fig.~\ref{sum_SE} shows the average spectral efficiency at different frequencies for indoor and outdoor scenarios as a function of the MIMO dimension. Increasing the MIMO dimension yields larger gains in indoor environments due to richer multipath propagation, particularly at lower frequencies, whereas outdoor environments benefit more at higher frequencies, reinforcing the need for site-specific, frequency-adaptive architectures.

\begin{figure*}[!t]
\centering



\begin{subfigure}[b]{0.48\textwidth}
  \centering
  \includegraphics[width=\linewidth]{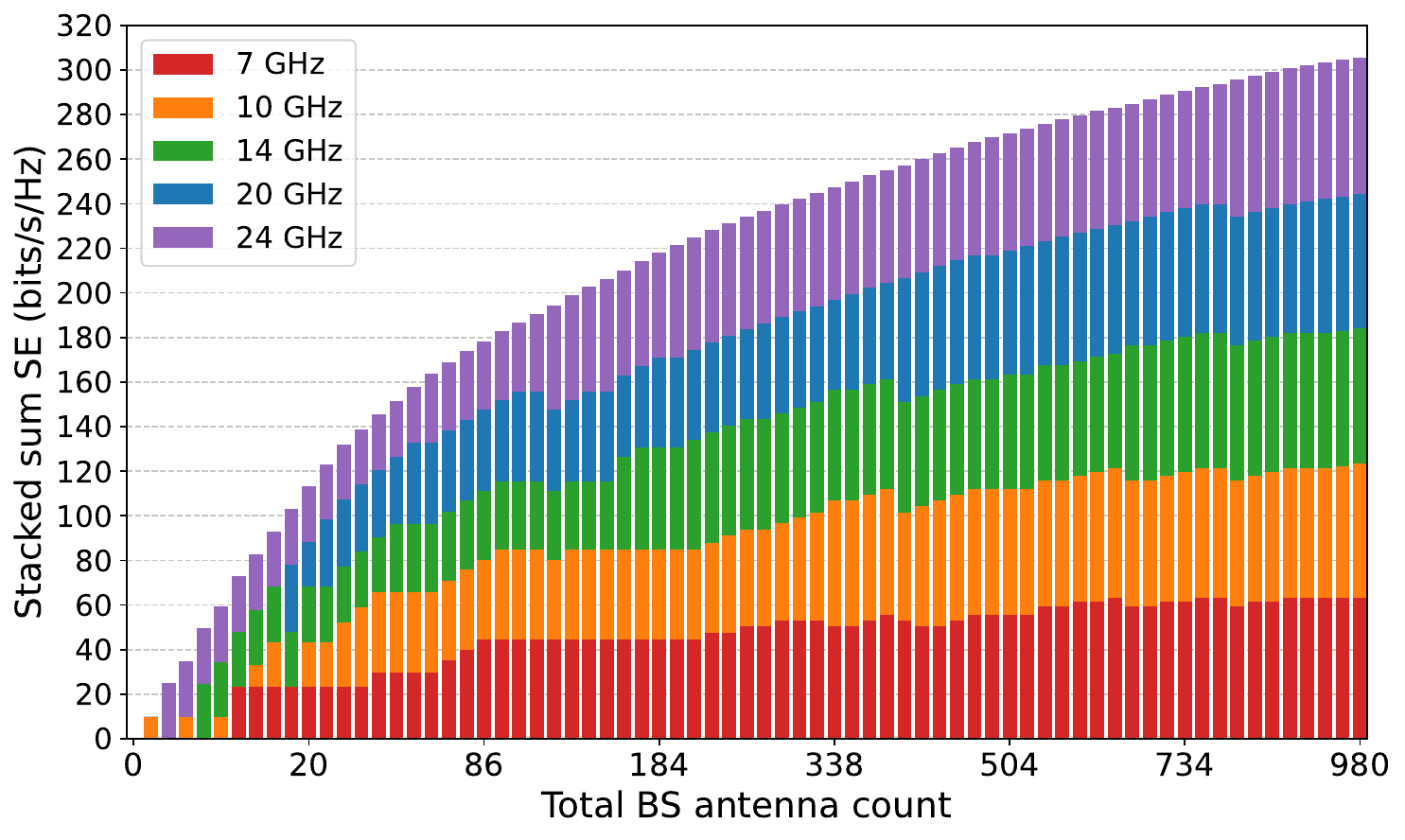}
  \caption{Indoor laboratory environment.}
  \label{stacked_SE_indoor}
\end{subfigure}
\hfill
\begin{subfigure}[b]{0.48\textwidth}
  \centering
  \includegraphics[width=\linewidth]{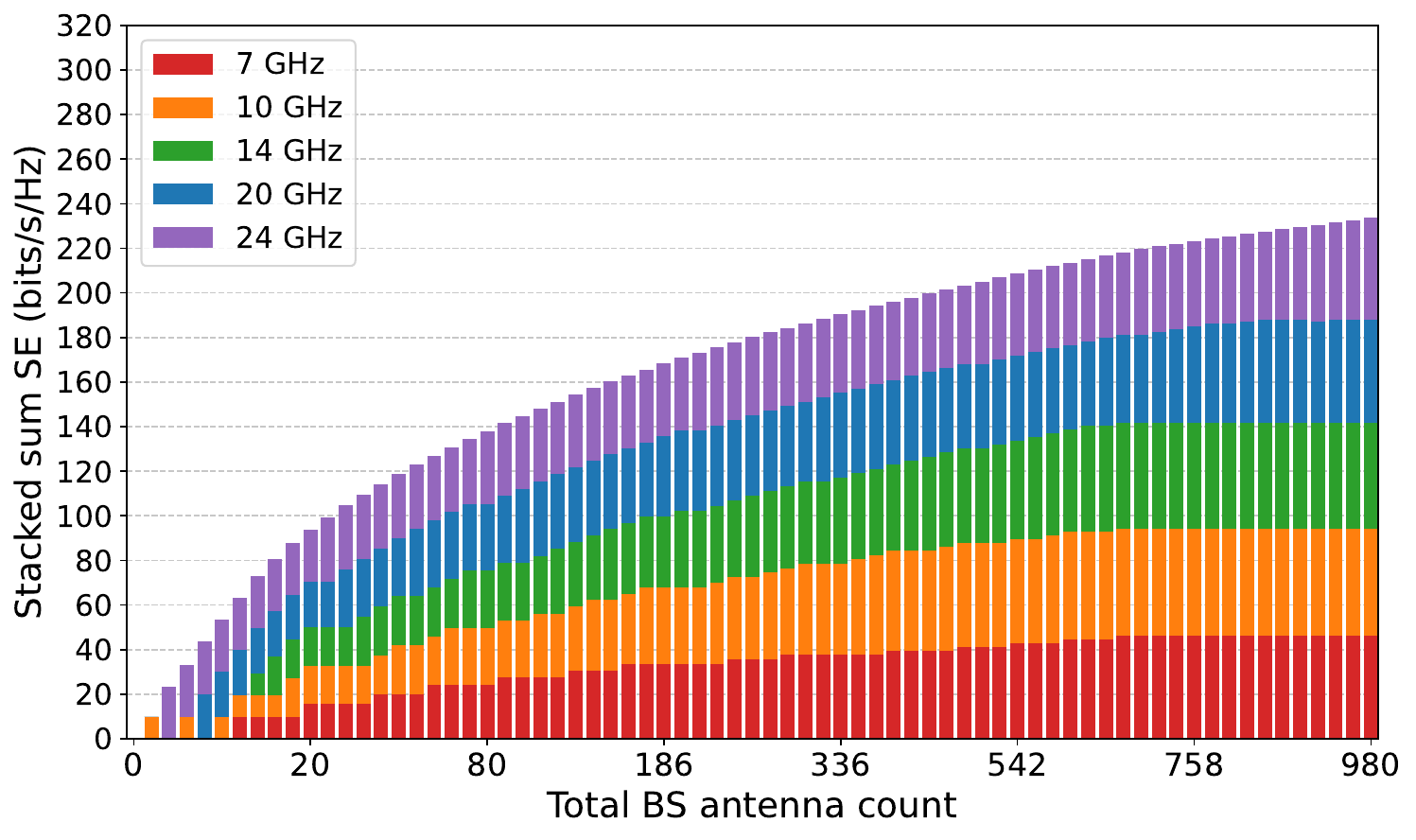}
  \caption{Outdoor urban macro environment.}
  \label{stacked_SE_outdoor}
\end{subfigure}

\caption{Stacked sum SE for the indoor and outdoor scenarios, for increasing frequency-adaptive ADC/DAC allocation optimum. \sofie{Already for low total Tx antenna counts, all bands are selected, which inducates that inter-band carrier aggregation is still preferred compared to MIMO, in case the bands are available.  Interestingly, all subbands have similar trends, and contribute proportionally, which means that also higher frequencies have significant MIMO multiplexing opportunities calling for large numbers of ADCs/DACs. Both indoor and outdoor channels have similar trends and frequency-dependent performance. The outdoor site shows clearly a lower peak stacked SE. } \sofie{X-axis should be total Rx antenna count!!!!! or basestation...}}
\label{stacked}
\vspace{-16pt}
\end{figure*}


In a second simulation scenario, we investigated hardware resource allocation as a function of the total number of available Base Station (BS) antennas for both indoor and outdoor scenarios, while keeping the User Equipment (UE) antenna \sofie{should be transmitter array} array fixed at 3×3. The BS antenna budget is varied from 0 to 980, corresponding to a maximum of a 14×14 array at all five frequencies. The resulting sum spectral efficiency is shown in Fig.~\ref{stacked}.

These XL-MIMO results confirm diminishing MIMO gains as antenna count increases, particularly in outdoor environments. In indoor deployments, MIMO gains are preferentially allocated to lower frequencies, while outdoor scenarios exhibit weaker frequency dependence. Spectrum aggregation is therefore favored even for relatively small antenna budgets.

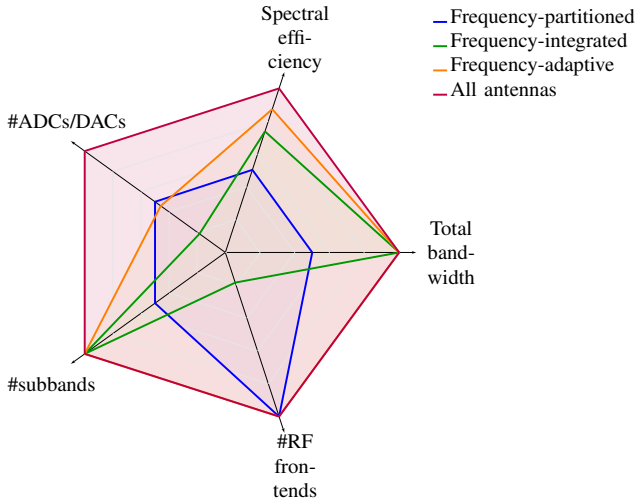
\begin{figure}[!t]
    \centering
    \resizebox{0.95\columnwidth}{!}{%
    \begin{tikzpicture}[
        every node/.style={font=\LARGE} 
    ]
    \begin{scope}[local bounding box=kiviat]
    \tkzKiviatDiagram[
        label distance=1cm,
        radial  = 5,
        gap     = 1,
        lattice = 5
    ]{Total bandwidth, Spectral efficiency,\#ADCs/DACs,\#subbands,\#RF frontends}

    \tkzKiviatLine[
        ultra thick,
        color=blue,
        fill=blue!20,
        opacity=.5
    ](2.5,2.518213,2.5,2.5,5)
    
    \tkzKiviatLine[
        ultra thick,
        color=green!60!black,
        fill=green!20,
        opacity=.5
    ](5,3.68886,0.918367,5,0.918367)
    
    \tkzKiviatLine[
        ultra thick,
        color=orange,
        fill=orange!20,
        opacity=.5
    ](5,4.36629,2.29592,5,5)
    
    \tkzKiviatLine[
        ultra thick,
        color=purple,
        fill=purple!20,
        opacity=.5
    ](5,5,5,5,5)
    
    
    \end{scope}
    
    \node[anchor=north west] at ([xshift=-20mm]kiviat.north east) {
    \setlength{\tabcolsep}{2pt} 
    \begin{tabular}{@{}l@{\hspace{2pt}}l@{}}
      \tikz\draw[blue,ultra thick] (0,0)--(0.4,0); & Frequency-partitioned \\
      \tikz\draw[green!60!black,ultra thick] (0,0)--(0.4,0); & Frequency-integrated \\
      \tikz\draw[orange,ultra thick] (0,0)--(0.4,0); & Frequency-adaptive \\
      \tikz\draw[purple,ultra thick] (0,0)--(0.4,0); & All antennas \\
    \end{tabular}
    };

    \end{tikzpicture}
    }
    
    \caption{Comparison of architectures in terms of design and performance parameters.}
    \label{spider4}
    \vspace{-12pt}

\end{figure}

Lastly, we analysed four different architectures
—frequency-partitioned, frequency-integrated, frequency-adaptive, and an idealized all-antennas baseline—across key performance and hardware parameters, in the outdoor environment. The comparison assumes that only the 7~GHz and 24~GHz subbands are available. The frequency-partitioned design 
has 5×196 antennas and RF frontends with 196 ADCs/DACs serving one subband at a time. The frequency-integrated architecture uses a 6×6 antenna array per subband with 36 dedicated ADCs/DACs each, totaling 180 ADCs/DACs. The frequency-adaptive architecture employs 5×196 antennas and frontends with 196 ADCs/DACs that can be dynamically assigned to a frontend belonging to any subband, as illustrated in Fig.~\ref{architecture}. The \textit{all-antennas} baseline uses 5×196 antennas and frontends with 5×196 dedicated ADCs/DACs. A visual representation of the performance comparison is shown in Fig.~\ref{spider4}. Both the frequency-adaptive and -integrated architectures use both available subbands, while the spectral efficiency of the frequency-adaptive approach is more than 18\% higher, illustrating the benefit of dynamically repurposing ADCs/DACs and baseband processing resources.



\balance

\section{Conclusion}\label{conclusion}

This paper investigates site-specific, frequency-dependent MIMO performance in 
FR3 using ray-tracing simulations in representative indoor and outdoor scenarios. The results demonstrate that FR3 supports 
MIMO operation with strong dependence on frequency and environment, motivating architectures that can adapt hardware resources across subbands. To this end, we propose a fully digital frequency-adaptive multi-band MIMO architecture that dynamically repurposes ADC/DAC and baseband processing resources in response to channel conditions and spectrum availability. Simulation results show that while exploiting additional spectrum is often optimal, frequency-adaptive resource allocation becomes particularly beneficial under partial spectrum availability and when MIMO gains concentrate at specific frequencies. Overall, the findings highlight the importance of site-specific, frequency-adaptive architectural design for efficient FR3 operation in future 6G systems.

\section*{Acknowledgement}
This work is partly supported by the MultiX project under the European Union’s Horizon Europe research and innovation programme (Grant No. 101192521). The work of Z. Cui was supported by the Research Foundation – Flanders (FWO), Senior Postdoctoral Fellowship under Grant No. 12AFN26N.

\balance
\bibliographystyle{IEEEtran}
\bibliography{Reference}

\end{document}